\documentclass{article}

\usepackage{PRIMEarxiv}
\usepackage{float}
\usepackage[utf8]{inputenc} 
\usepackage[T1]{fontenc}    
\usepackage{hyperref}       
\usepackage{url}            
\usepackage{booktabs}       
\usepackage{amsfonts}       
\usepackage{nicefrac}       
\usepackage{microtype}      
\usepackage{lipsum}
\usepackage{fancyhdr}       
\usepackage{graphicx}       
\usepackage{footmisc}
\usepackage[english]{babel}
\usepackage{ulem}
\usepackage{booktabs}
\usepackage{multirow}

\addto\captionsenglish{}
\graphicspath{{media/}}     

\title{ Language-Inspired Modeling Reveals Redundant Encoding of N4-acetylcytidine(ac4C) Modifications in mRNA
}
\author{
  Li Yang, Dongbo Wang\textsuperscript{*} \\
  School of Information Management, Nanjing Agricultural University, Nanjing, 210095, China \\
}

\begin{document}
\maketitle
\setlength{\footnotemargin}{0pt} 
\renewcommand{\thefootnote}{}     
\footnote{*Email: db.wang@njau.edu.cn (D. Wang)}
\begin{abstract}
The ac4C modification on mRNA has been demonstrated to be associated with various diseases; however, its molecular mechanism remains unclear. The wet lab experiments produced relatively rough data, which lack precise ac4C modification sites, and extracting valuable information from such data remains a challenge. In this study, we integrate linguistics, traditional machine learning, and deep learning, establishing a link between the understanding of mRNA data and natural language processing (NLP). Through our analysis, we successfully revealed key information about ac4C in mRNA and uncovered the information storage mechanism of ac4C redundancy on a single sequence. This redundant information storage method in mRNA facilitates the transmission of ac4C information and promotes the enrichment of ac4C.
\end{abstract}


\section{Introduction}

 ac4C is a newly identified epitranscriptional modification catalyzed by the N-acetyltransferase NAT10. Studies have demonstrated that ac4C enhances RNA stability and translation efficiency, thereby promoting protein synthesis\cite{1}, while its aberrant modification is closely associated with various pathological processes, including cancer and immune diseases\cite{2}. Therefore, elucidating the mechanism of ac4C modification will facilitate the development of targeted therapies aimed at either correcting or leveraging these modifications for disease treatment\cite{3}. However, the underlying mechanism of ac4C modification remains poorly understood.
 Although the role of ac4C in regulating RNA stability has been preliminarily validated, its detection remains challenging due to its sensitivity to environmental factors such as temperature, pH, and oxidative stress\cite{4}. Currently, no single method can simultaneously achieve high sensitivity, single-nucleotide resolution, and high-throughput detection, which significantly limits the comprehensive study of ac4C modifications. ac4C RIP-seq (RNA Immunoprecipitation Sequencing) has become one of the most commonly used techniques for identifying ac4C modifications. This method enables the capture of ac4C modifications at the mRNA fragment level\cite{5} and has provided crucial insights into the potential functions of ac4C in cell proliferation, differentiation, and cancer progression\cite{6,7,8,9}. However, a major limitation of this approach is its inability to precisely pinpoint single-nucleotide modification sites, thereby restricting a deeper understanding of the mechanisms underlying ac4C function. To enhance detection accuracy, researchers have incorporated peak calling into ac4C RIP-seq data analysis and rigorously filtered data to remove artifacts such as non-specific binding to immunoglobulin (IgG) and replicability1. A thorough analysis of these data is expected to elucidate the recognition mechanisms of ac4C modifications on mRNA, thereby advancing our understanding of its biological functions.
 
 In our previous research, we investigated DNA information encoding from a linguistic perspective, integrating DNA dataset cleaning and information storage with natural language processing (NLP). Through this approach, we successfully elucidated the information storage mechanism of N6-adenine (6mA) methylation in DNA \cite{10} and its search mechanism\cite{11}. RNA and DNA exhibit similarities in symbolic representation, with RNA composed of four nucleotides—A, U, C, and G—and existing as a single-stranded molecule in the cell. Building upon these similarities, this study seeks to elucidate the information storage mechanism of ac4C in mRNA from a linguistic perspective. Unlike 6mA methylation in DNA, the precise nucleotide positions of ac4C modifications remain unknown, with only relative peak positions available. In the absence of specific recognition sequences, we will attempt to integrate a linguistic approach with deep learning to comprehensively analyze ac4C modifications in mRNA. Our results demonstrate conserved motif patterns across multiple datasets and indicate that ac4C modifications exhibit redundancy within individual mRNA transcripts. This redundancy may contribute to the stability of ac4C-mediated information transmission in dynamic environments. Finally, based on relative peak position data, we successfully identified key ac4C modification patterns and their redundant information storage properties in mRNA. These findings shed light on the interaction between NAT10 and mRNA and provide a novel framework for understanding the molecular mechanisms of ac4C modifications.
\section{Methods}
\paragraph{Data Collection and Preprocessing}
Arango et al.\cite{1} initially identified 4,250 candidate ac4C peaks using acRIP-seq techniques. However, since acRIP-seq is not a base-resolution method, these ac4C sites are not necessarily cytidines, meaning that the precise ac4C modification sites remain unknown. To define peak regions and peak summits, Arango et al. subsequently performed peak calling and implemented stricter filtering criteria, reducing the dataset to 2,153 high-confidence ac4C peaks. These refined peaks were considered more reliable. In this study, we selected 2,153 ac4C peaks to construct the dataset. To establish a relatively reliable dataset, we defined ac4C peaks and their adjacent cytidines as modification sites. Centered on these modification sites, we extracted 25 bp upstream and 25 bp downstream from the peak, resulting in 51-bp RNA fragments.To mitigate the impact of redundant sequences, we applied CD-HIT\cite{12} with a similarity threshold of 0.8 to eliminate highly similar sequences, yielding 1,249 positive samples. Negative samples were randomly selected from transcripts that do not contain ac4C modifications. These transcript sequences were obtained from the UCSC Genome Browser, and 51-bp fragments centered on cytidines were extracted, ultimately yielding 1,249 negative samples. Arango et al.¹ demonstrated that most mRNAs contain one to two ac4C peaks. In addition to peak data, subpeak data were also identified. To construct a comparative dataset, we applied the same method to collect ac4C data from secondary peaks.Finally, we obtained two sets of mRNA ac4C datasets, as shown in Table 1.
\begin{table}[ht]
	\centering
	\begin{tabular}{|l|l|l|}
		\hline
		Dataset	& Positive Samples	& Negative Samples \\
		\hline
		Peak dataset & 1,249 & 1,249 \\
		\hline
		Subpeak dataset & 1,302	& 1,302 \\
		\hline
	\end{tabular}
	\caption{\label{tab1}Two independent ac4C datasets 1.0}
\end{table}

\paragraph{Preliminary Exploration of Sequence Characteristics of ac4C on mRNA}

To investigate the nucleotide distribution around ac4C peaks, we utilized the online tool MEME\cite{13} to identify conserved motifs in sequences surrounding ac4C modification sites. As shown in Figure 1, sequences containing ac4C modifications exhibit a highly enriched motif (CXX), indicating a C-rich subsequence, which is consistent with previous observations.

\paragraph{Model Selection and Training}

 Convolutional Neural Network (CNN)\cite{14} were originally designed for image data but have been widely applied in the field of Natural Language Processing (NLP). Particularly in sentence or text processing, CNNs can effectively capture local features such as key phrases and contextual relationships. Therefore, in this study, we selected CNN as the testing model. The dataset was split into training and test sets at a ratio of 8:2, ensuring no overlap between them. Additionally, we performed 5-fold cross-validation on the training data to enhance the model's robustness and generalization ability.

\paragraph{Performance evaluation}

We evaluated the model's performance using accuracy (ACC), precision (P), recall (R), F1-score, Matthews correlation coefficient (MCC), and the area under the receiver operating characteristic (ROC) curve (AUC). The definitions of these metrics are as follows:
\begin{equation}
	P = \frac{TP}{TP + FP}
\end{equation}

\begin{equation}
	R = \frac{TP}{TP + FN}
\end{equation}

\begin{equation}
	\frac{1}{F_1} = \frac{1}{2} \left( \frac{1}{P} + \frac{1}{R} \right)
\end{equation}

\begin{equation}
	ACC = \frac{TP + TN}{TP + TN + FN + FP}
\end{equation}

\begin{equation}
	MCC = \frac{TP \cdot TN - FP \cdot FN}{\sqrt{(TP + FN)(TP + FP)(TN + FP)(TN + FN)}}
\end{equation}
Here, TP (true positives) and TN (true negatives) represent correctly classified positive and negative samples, respectively, while FP (false positives) and FN (false negatives) denote misclassified samples.

\paragraph{Simulating Base Transition States with Chinese Characters}

As the 4 Features description was not changing the DNA sequence, it means that all DNA In our previous study\cite{10}, we utilized a manually constructed dictionary to map four-character DNA sequences onto the Chinese language space, simulating the base transition states during catalytic reactions. However, the feasibility of this approach in RNA has not yet been validated. To address this, the present study applies the same dictionary-based mapping approach to RNA sequences, representing four-character RNA sequences in the Chinese language space to simulate base transition states during mRNA acetylation reactions. Furthermore, we evaluate its impact on ac4C-related information in mRNA. We use two types of feature representations: four symbolic features and 204 (4×51) features transformed into the Chinese language space. These features are then input into a CNN. sequences only have four features(A, T, C, and G). 
\paragraph{The Impact of Different Linguistic Features on ac4C Information}

We found that the Chinese character-based simulation did not introduce significant interference, indicating that simulating the base transition states of mRNA during the ac4C reaction phase using Chinese characters is feasible. To investigate the impact of different Chinese character feature representations on ac4C-related information in mRNA, we applied three different mapping dictionaries (Dictionary 1(Dict1), Dictionary 2(Dict2), and Dictionary 3(Dict3)) for transformation. The features obtained from each dictionary were separately input into a CNN model, followed by 5-fold cross-validation and evaluation. The results are shown in Fig. 3.

\paragraph{Rule Extraction (Motif Mining)}

We employed the Apriori \cite{15} for rule mining, setting the support threshold to 0.06 and the confidence threshold to 1.0. The process was as follows: first, the positive samples from the two datasets were converted into Chinese character representations using Dictionary 1. Then, the Apriori algorithm was applied to extract association rules from the positive samples. These rules suggest that the occurrence of these key nucleotides is associated with ac4C modification at the central C.

\paragraph{Motif Distribution}
To investigate the characteristic distribution of motifs in the samples, we calculated the proportions of these motifs in both positive and negative samples, which reflects their occurrence tendency across different categories. Notably, the negative samples were randomly generated, leading to differences between the peak dataset and the sub-peak dataset. The calculation formula is as follows:
\begin{equation}
	P(pos) = \frac{N(pos)}{Npos) + N(neg)}
\end{equation}
Where:

\setlength{\parindent}{2em} •	P(pos) represents the proportion of a motif in positive samples.

\setlength{\parindent}{2em} •	N(pos) indicates the  number of occurrences of a motif in positive samples.

\setlength{\parindent}{2em} •	N(neg) indicates the  number of occurrences of a motif in negative samples.

If P(pos) approaches 1, it suggests that the motif appears almost exclusively in positive samples; if it is close to 0.5, the motif is evenly distributed between positive and negative samples; and if it is close to 0, the motif is predominantly found in negative samples.
Despite rigorous filtering, including peak calling and artifact removal, the peak and sub-peak positive samples do not necessarily represent true ac4C modification sites. However, we can be certain that these positive samples contain true ac4C modification sites, even though their exact quantity remains unknown. The peak dataset and the sub-peak dataset are two completely independent datasets. However, it is important to recognize that ac4C modification signals should be consistent between these datasets and exhibit similar patterns, which is precisely the information we aim to uncover. Surprisingly, after computational analysis and ranking, as shown in the supplementary table, we identified a set of structurally similar motifs composed of C and G. These motifs exhibit a consistent structural composition, differing only slightly at individual nucleotide positions. Drawing from our previous DNA data analyses, we recognized that these motifs are likely NAT10-recognized signals on mRNA, which will be further validated in subsequent experiments. Additionally, to examine the distribution of motifs along a single mRNA sequence, we selected motifs with an occurrence proportion of 90\% to annotate individual mRNA sequences, as illustrated in Supplementary Table and Figure 4.

\paragraph{False-Negative Sample Gradient Cleaning Experiment}

To determine whether the aforementioned motifs constitute the information that NAT10 recognizes on mRNA, we set P(pos) = 0.90, 0.85, and 0.80 as filtering thresholds to identify potential false-negative samples. If a motif appears in negative samples and its P(pos) exceeds the corresponding threshold, those negative samples are classified as false negatives. We then exclude these negative samples containing such motifs and re-evaluate the model’s classification performance to assess whether these motifs act as key indicators of ac4C modification.

\paragraph{Effect of a Cleaner ac4C Dataset on CNN Model Performance}

To further validate the aforementioned motifs, we set P(pos) = 0.85 (Supplementary Tables 1 and 2) as the filtering threshold. We then randomly selected an equal number of negative samples from transcripts of different species, ensuring that these negative samples no longer contained such motifs. By combining these newly selected negative samples with the previously filtered positive samples, we constructed a new dataset, referred to as dataset 0.85. In this dataset, the number of positive and negative samples remained equal; however, the negative samples were filtered to remove false negatives based on the predefined threshold. This dataset refinement process resulted in a more curated dataset. Compared with the original, unfiltered Dataset-1.0, the classification performance of the CNN model was expected to reflect the significance of these motifs in ac4C modification.

\paragraph{Comparative Study of Different Deep Learning Models}
To evaluate the learning capabilities of different types of deep learning models on mRNA data, we introduced a lightweight traditional deep learning model (LSTM\cite{16}) and a comprehension-based large language model (BERT\cite{17}) for comparison on the newly constructed Dataset-0.85. All models used four symbolic representations as input and were trained and evaluated using five-fold cross-validation to ensure the robustness of the experimental results.

\section{The Sequence Characteristics of ac4C in mRNA }

\begin{figure}[H]
	\centering
	\includegraphics[width=0.8\textwidth]{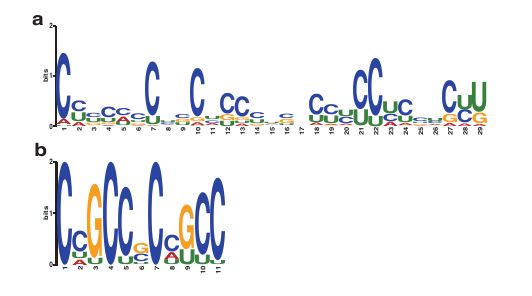}
	\caption{\textbf{Preliminary Exploratory Analysis Results of ac4C-Modified Sequences.} }
	\label{fig1} 
\end{figure}

To examine the nucleotide distribution of the collected ac4C-modified positive samples, we performed motif analysis using the MEME software, as shown in Fig. \ref{fig1}a and Fig. \ref{fig1}b. The ac4C-modified sequences were strongly enriched in the motif (CXX), which is consistent with previous observations\cite{1,18,19}. These findings provide a strong basis for subsequent analyses.

\section{Simulating Base Transition States with Chinese Characters }

\begin{figure}[H]
	\centering
	\includegraphics[width=\textwidth]{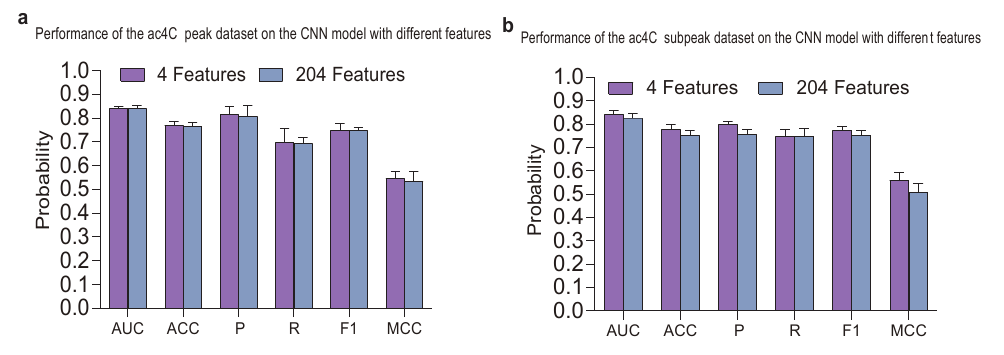}
	\caption{ \textbf{Performance comparison of CNN models with 4-feature and 204-feature inputs in terms of AUC, ACC, P, R, F1, and MCC metrics.}} 
	\label{fig2} 
\end{figure}

Based on our experience in DNA sequence data analysis, we similarly hypothesized that the representation of bases at different positions on mRNA should vary during the ac4C reaction process. By using Custom Dict1, we mapped RNA to the Chinese language, which expands the original four-feature symbolic representation into a sequence with 204 feature representations. Two datasets with different representations before and after the transformation were then fed into a CNN model to evaluate the impact of feature conversion on ac4C information within the sequence. As shown in Fig. \ref{fig2}, across multiple evaluation metrics on the two collected datasets, we observed that, apart from a slightly larger difference in MCC and P values in Fig. \ref{fig2}b, the differences across other evaluation metrics were minimal. This suggests that mapping RNA into the human language space appears to be a feasible approach.
\section{The Influence of Different Linguistic Features on ac4C Modification Information}

\begin{figure}[H]
	\centering
	\includegraphics[width=\textwidth]{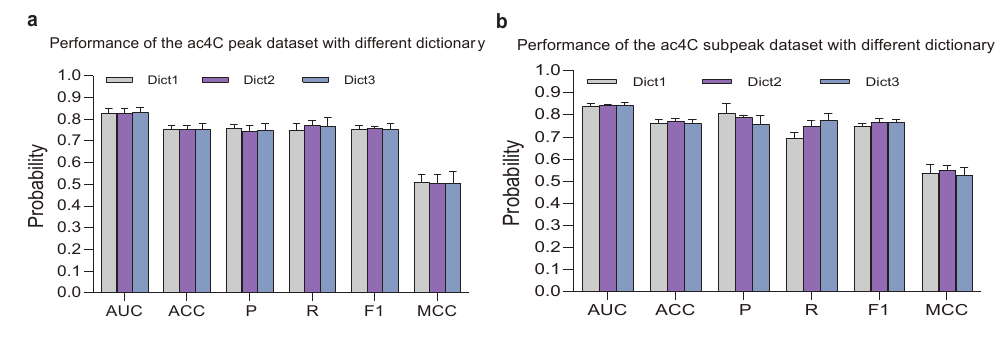}
	\caption{\textbf{Comparison of CNN models using different dictionaries, focusing on AUC, ACC, P, R, F1, and MCC.}
	} 
	\label{fig3} 
\end{figure}

After testing the aforementioned dictionary conversion schemes, we further analyzed the impact of different dictionary mappings on ac4C information in mRNA. As shown in Fig. \ref{fig3}, across the two datasets, the overall differences among different mapping dictionaries were not significant, except for some variations in P and R values in Fig. \ref{fig3}b. Therefore, we can reasonably conclude that different dictionary mapping schemes do not have a significant impact on ac4C information in mRNA. 
\section{Key Information and Storage Patterns of ac4C Modifications on mRNA}
\begin{figure}[H]
	\centering
	\includegraphics[width=\textwidth]{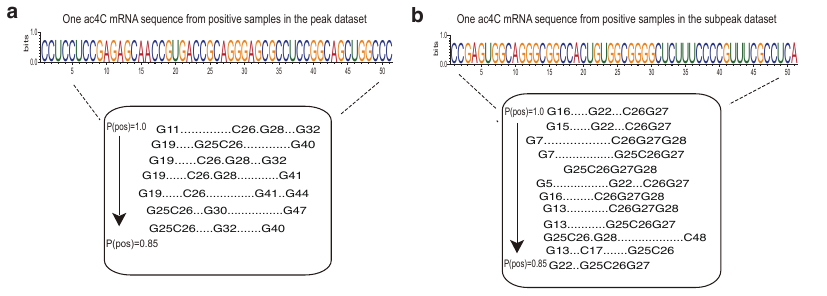}
	\caption{\textbf{Illustration of the storage pattern of ac4C modification information with P(pos) $\geq 0.85$  in a single mRNA sequence.}
	} 
	\label{fig4} 
\end{figure}

In this study, we mapped RNA data represented by four symbols into a multi-feature representation in the Chinese language and conducted a comparative analysis using a CNN model. The results demonstrate that mapping RNA language to natural language (Chinese) is feasible. Furthermore, we applied the Apriori rule mining algorithm to analyze patterns in two datasets to extract key features of ac4C modification. The results indicate that the mined rules from both datasets are highly similar, with the dominant motif consisting of three guanine nucleotides (G) and one cytosine nucleotide (C) (see Fig. \ref{fig4} and Supplementary Tables 1 and 2). In Fig. \ref{fig4}b, the G\uline{C}GG motif in mRNA (with the underlined position marking the ac4C modification site) was also observed in DNA’s N6-methyladenine (6mA) epigenetic modification, where a similar motif, specifically G\uline{A}GG (with the underlined position marking the 6mA modification site), was identified\cite{11}. Additionally, we observed that in mRNA, the distribution pattern of this motif is relatively dispersed, a feature that aligns with the dynamic characteristics of mRNA as a single-stranded molecule. To further verify the stability of the rules, we extended the original data by adding two additional bases on both sides of the peak and sub-peak regions, constructing two new ac4C datasets and conducting the same rule mining analysis. The results show that all four datasets consistently yield a similar pattern consisting of three G and one C, further supporting the importance of this motif in ac4C modification (see Supplementary Tables 1, 2, 3, and 4).

As shown in Fig. \ref{fig4}, applying these rules (P(pos) $\geq 0.85$) to individual sequences has revealed redundancy in ac4C information storage. Furthermore, multiple dispersed motifs in mRNA can collectively form "binding hotspots," allowing proteins to bind to multiple motifs rather than being restricted to a single site. This binding pattern may lead to the local enrichment of ac4C modifications rather than their uniform distribution across the entire sequence\cite{1}. These findings suggest that motif-mediated multi-site binding could be a direct cause of the fragmented distribution of ac4C modifications.
\section{Gradient Washing Experiment on the Dataset}

\begin{figure}[H]
	\centering
	\includegraphics[width=\textwidth]{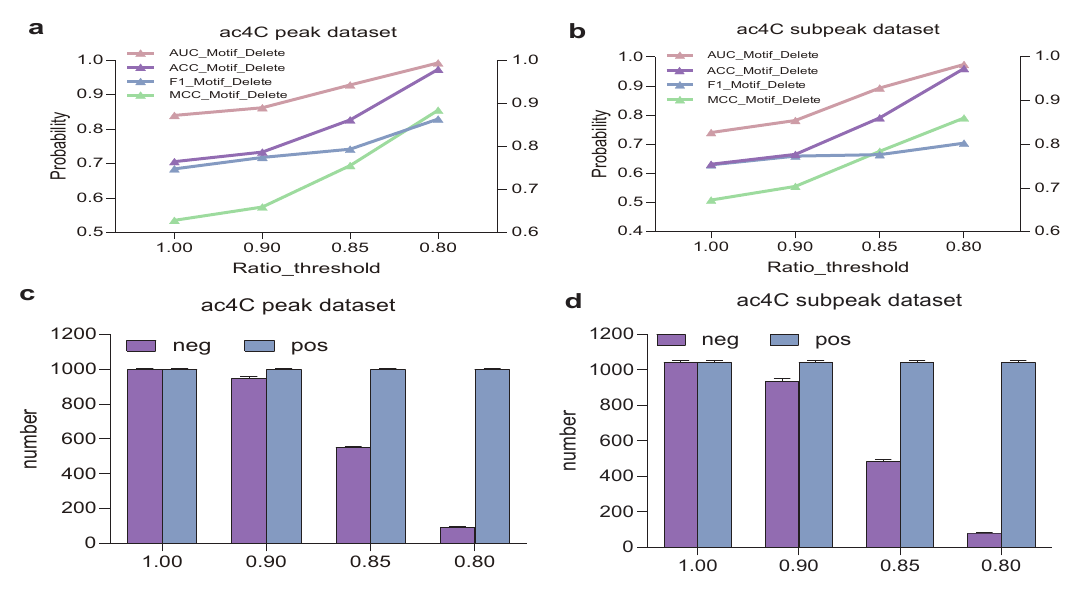}
	\caption{\textbf{Evaluation of Key Information for ac4C through Data Cleaning by Removing False Negative Samples.} 
		a, b. False negative samples are removed based on P(pos) thresholds of 0.90, 0.85, and 0.80, and the CNN model’s performance is evaluated using AUC, ACC, F1, and MCC across two independent datasets. b, c. Changes in the number of positive and negative samples after applying different thresholds (0.90, 0.85, 0.80) for negative sample removal.
	} 
	\label{fig5} 
\end{figure}

To further validate whether the motifs identified through rule mining serve as key information for ac4C modification in mRNA, we set different false-negative determination thresholds, where the proportion of motifs in positive samples (P(pos)) was set to 0.90, 0.85, and 0.80, respectively, as the data cleaning process intensified. Specifically, we removed negative samples containing these motifs to identify potential noise or incorrectly labeled data, thereby assessing the reliability of ac4C modification information. The experimental results showed that after removing negative samples containing the motifs, the CNN model exhibited a steady improvement in its classification performance, with increases observed in AUC, ACC, F1, and MCC scores (Fig. \ref{fig5}a and Fig. \ref{fig5}b). 
This further validated the significance of the extracted rule-based information in ac4C modification recognition. Additionally, as shown in Fig. \ref{fig5}c and Fig. \ref{fig5}d, we observed changes in the proportion of these rules within the dataset, further supporting their critical role in ac4C modification.

\section{Impact of a Cleaner ac4C Dataset on the CNN Model }

Building upon previous gradient washing experiments, we set P(pos) = 0.85 as the false-negative threshold. Negative samples lacking these motifs were randomly selected from transcripts without ac4C modifications, producing a cleaner dataset (Dataset 0.85) with an equal number of positive and negative samples. With this new dataset, we conducted two sets of experiments. In the first experiment, we tested different feature representations as model inputs and found that the CNN model achieved better performance when using the more expressive Chinese character representation, compared to Fig. \ref{fig2}. In the second experiment, we compared Dataset 0.85 to the unfiltered original dataset (Dataset 1.0). As shown in Figures c and d, after additional data cleaning, the CNN model exhibited improvements across multiple evaluation metrics, including AUC, ACC, P, and R. These results suggest that these motifs play a crucial role in data cleaning and further validate their significance as key features in ac4C recognition.
\begin{figure}[H]
	\centering
	\includegraphics[width=1\textwidth]{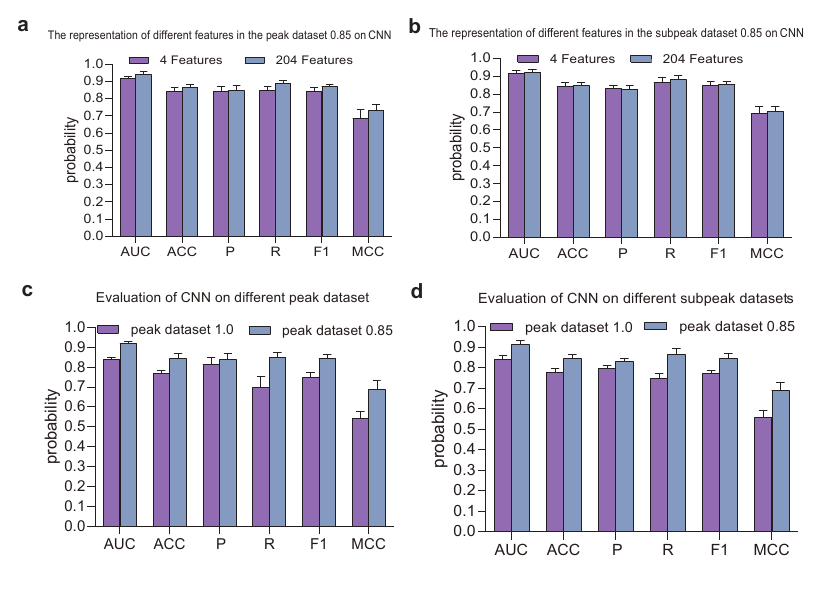}
	\caption{\textbf{ The Impact of a Cleaner ac4C Dataset on CNN Model Performance.} 
		a, b Performance comparison of CNN models using 4 and 204 features as input on the new dataset 0.85, evaluated across AUC, ACC, P, R, F1, and MCC metrics. c, d  Performance differences of CNN models in terms of AUC, ACC, P, R, F1, and MCC between the 0.85 dataset after cleaning and the raw 1.0 dataset.
	} 
	\label{fig6} 
\end{figure}

\section{Comparison of Different Models}

To evaluate the ability of different deep learning models to learn from the ac4C dataset, we selected models commonly used in natural language processing, including traditional deep learning models (CNN and LSTM) as well as a pretrained language model (BERT). As shown in Table \ref{tab:table2}, the pretrained language model BERT outperformed traditional deep learning models in learning ac4C modification information, achieving higher scores across all evaluation metrics (AUC, ACC, P, R, F1, and MCC).

\begin{table}[H]
	\caption{\textbf{The performance of different models on the Peak dataset at 0.85 and the Subpeak dataset at 0.85}}                       
	\centering 
	\label{tab:table2}                                      
	\begin{tabular}{ccccccccccccc}                         
		\toprule                                     
		\multirow{2}{*}{Method}&                      
		\multicolumn{6}{c}{Peak dataset 0.85}&            
		\multicolumn{6}{c}{Subpeak dataset 0.85}\cr     
		\cmidrule{2-7}\cmidrule{8-13}                 
		& AUC&ACC&P&R&F1&MCC&AUC&ACC&P&R&F1&MCC\\
		\midrule                                      
		LSTM &0.929	&0.858&	0.849&	0.872&	0.860&	0.717&	0.904&	0.839&	0.819&	0.871&	0.844&	0.680\\
		CNN	&0.918&	0.844&	0.841&	0.850&	0.845&	0.689&	0.916&	0.846&	0.831&	0.867&	0.849&	0.692\\
		Bert&0.947&	0.873&	0.891&	0.852&	0.870&	0.749&	0.929&	0.860&	0.880&	0.834&	0.856&	0.722\\
		\bottomrule                                   
	\end{tabular}
\end{table}

\section{Discussion}

In this study, we draw an analogy between mRNA and human language (Chinese), combining RNA dataset cleaning with natural language understanding to validate the feasibility of converting mRNA into Chinese. Next, we applied a rule-mining approach to extract rules from a fuzzy mRNA dataset without precise ac4C modification sites, and validated these rules through multiple datasets. The results indicate that the ac4C modification features on mRNA exhibit a tendency to regress to motifs with similar combinations, primarily consisting of 3 G and 1 C. We further employed deep learning techniques to learn high-dimensional features and repeatedly validated these key pieces of information. Experimental results show that these motifs indeed represent critical information regarding ac4C modification on mRNA. Finally, we found that ac4C modification information in a single mRNA sequence is redundant, suggesting that mRNA may store key information through multiple similar motifs. This redundancy likely enhances the stable transmission of ac4C information in a liquid environment.

The redundant presence of loosely organized motifs on mRNA creates favorable conditions for the cooperative action of multiple proteins in RNA cytosine acetylation. Recent low-temperature electron microscopy studies have shown that the NAT10 protein forms a symmetric heart-shaped dimer during RNA cytosine acetylation\cite{20}, a structure that supports the idea of multi-protein cooperation. The study also found that the surface of NAT10 contains other basic patches that can participate in non-specific RNA binding, which can compensate for the loss of RNA binding affinity at specific binding sites\cite{20}. This aligns with the loosely organized state observed between the ac4C motif elements on mRNA. At the same time, these dispersed 3 G can form more hydrogen bonds, providing strong support for stabilizing highly dynamic RNA structures. Furthermore, many RNA-binding proteins have a modular structure composed of several small domains with repeated sequences, which allows recognition of long sequence segments and separated sequences\cite{21}. These modular arrangements can coordinate and enhance binding with RNA.

Multiple similar motifs are present on mRNA, which can increase the probability of RNA-binding protein (RBP) binding. These multiple loosely organized motifs may form "binding hotspots," allowing proteins to choose the optimal binding state from multiple possible conformations without requiring significant structural adjustments. This binding pattern can also influence downstream translational regulation mechanisms, such as enhancing the probability of specific RBP binding, affecting mRNA fate (stability, translation efficiency); and enabling dynamic regulation, where different motifs may become the dominant binding sites under different environmental conditions.

Through interdisciplinary integration, we have addressed the issue of extracting ac4C information from mRNA, which lacks accurate modification sites, thus overcoming the limitations of existing technologies. Ultimately, this study reveals the characteristic patterns of ac4C modifications on mRNA and their redundant storage mechanisms, providing important insights for a deeper understanding of the functional regulation.

\setlength{\parindent}{0pt}This paper has supplementary information.

\setlength{\parindent}{0pt}Correspondence and material requests should be addressed to db.wang@njau.edu.cn .

\section*{Data and Code availability}

The data that support the findings of this study are available from GitHub. The GitHub repository will be made publicly available upon formal publication.

\section*{Acknowledgments}
This work was supported by the Major Project of the National Social Science Fund of China (Project No. 21\&ZD331), which enabled the development of this research. We would like to express our sincere gratitude to the funding agency for their trust and support. We also thank all colleagues for their invaluable expertise and assistance throughout the course of this project.

\bibliographystyle{unsrt}  
\bibliography{references}

\end{document}